\PassOptionsToPackage{british}{babel}
\documentclass[fleqn,10pt]{wlscirep}
\usepackage[utf8]{inputenc}
\usepackage[T1]{fontenc}
\usepackage{lineno}
\usepackage{multirow}
\usepackage{booktabs, tabularx}
\usepackage{amssymb} % For math symbols
\usepackage{textcomp}
\usepackage[british]{babel} % Use British English spelling
\usepackage{graphicx}

\usepackage{hyperref}
\usepackage{caption}
\usepackage{subcaption}
\usepackage{tikz}
\usepackage{float}

\usepackage{xcolor}

\title{Cosmos 1.0: a multidimensional map of the emerging technology frontier}

\author[1,$\dag$]{Xian Gong}
\author[1,2,$\dag$]{Paul X.~McCarthy}
\author[3]{Colin Griffith}
\author[1]{Claire McFarland}
\author[1]{Marian-Andrei Rizoiu}

\affil[1]{University of Technology Sydney, Faculty of Engineering and Information Technology, Sydney, 2007, Australia}
\affil[2]{University of New South Wales, Sydney, 2052, Australia}
\affil[3]{Data61, Commonwealth Scientific and Industrial Research Organisation (CSIRO) Australia}

\affil[*]{corresponding author(s): Xian Gong (xian.gong@student.uts.edu.au)}

\affil[$\dag$]{these authors contributed equally to this work}

\begin{abstract}

This paper introduces the Cosmos 1.0 dataset and describes a novel methodology for creating and mapping a universe of technologies, adjacent concepts, and entities. We utilise various source data that contain a rich diversity and breadth of contemporary knowledge. The Cosmos 1.0 dataset comprises 23,544 technology-adjacent entities (TA23k) with a hierarchical structure and eight categories of external indices. Each entity is represented by a 100-dimensional contextual embedding vector, which we use to assign it to seven thematic tech-clusters (TC7) and three meta tech-clusters (TC3). We manually verify 100 emerging technologies (ET100). This dataset is enriched with additional indices specifically developed to assess the landscape of emerging technologies, including the Technology Awareness Index, Generality Index, Deeptech, and Age of Tech Index. The dataset incorporates extensive metadata sourced from Wikipedia and linked data from third-party sources such as Crunchbase, Google Books, OpenAlex and Google Scholar, which are used to validate the relevance and accuracy of the constructed indices.

\end{abstract}

\begin{document}

\flushbottom
\maketitle
%  Click the title above to edit the author's information and abstract
\thispagestyle{empty}

% \noindent Please note: Abbreviations should be introduced at the first mention in the main text – no abbreviations lists or tables should be included. The structure of the main text is provided below.

\section*{Background \& Summary}

Emerging technologies have been shown to deliver the greatest benefits to the economy and society when understood and adopted early, resulting in improved health, environmental outcomes, economic growth, and sustained innovation~\cite{balakrishnan1979discussion, martin1995foresight}. In the era of the Fourth and even Fifth Industrial Revolution (Industry 4.0 and 5.0), researchers are delving deeply into defining what constitutes emerging technologies~\cite{rotolo2015emerging, alvarez2021technological}. These technologies may be entirely new innovations or reimagined applications of existing technologies. Their critical role in national competitiveness~\cite{dahlman2007technology, coccia2019nations} and a firm’s innovative growth~\cite{lee2004strategic} is increasingly evident. We can see this with the digital transformation of the retail industry. Traditional store-based retailers who adopted the Internet as a communication channel and formed e-alliances experienced a positive impact on firm performance. This is supported by a study reviewing 181 companies across multiple countries, demonstrating the broad impact of digital adoption in the retail sector~\cite{vaishnav2023thematic}. Technologies related to artificial intelligence, smart devices, information and communication technologies, new materials, robotics, automation, sensors, and mechatronics, among others, are proving to be pivotal elements in the competitiveness of both developed and emerging economies~\cite{alvarez2021technological}. 

Both qualitative and quantitative methods are currently used to identify emerging technologies. Traditionally, the most common way is to use qualitative "top-down" processes with panels of experts discussing and voting on critical technologies and themes. Many leading global organisations use this approach, for example, the Organisation for Economic Co-operation and Development (OECD), World Economic Forum (WEF), and MIT Technology Review each produce annual reports on emerging technologies~\cite{/content/publication/sti_in_outlook-2016-en, WEF2023, MITTechReview2023}. The best-typified one is the Delphi method, a structured methodology designed to systematically elicit and refine the opinions of experts through iterative rounds of questionnaires and controlled feedback~\cite{brown1968delphi}. Quantitative methods evolve rapidly with text analysis and deep learning techniques. There are two main limitations in current methods of identifying emerging technologies: the types of data used and the "top-down" methodology applied. At an early stage, most data used to explore emerging technologies started with publications, patent information, News articles or a mix of these~\cite{abercrombie2012study, zhang2014term}. Counts of keywords, authors, publications, citations, patents, or even News articles are commonly used as data resources of quantitative methods~\cite{bettencourt2008population, ho2014technological}. The development of data analysis tools and methods, including natural language processing (NLP), subject-action-object (SAO) structure, similarity matrix, knowledge networks, full-text analysis, and large language models (LLMs), has the potential to enable the creation of new types of data for detecting emerging technologies~\cite{kim2020sao2vec, bonaccorsi2020emerging}. The vast majority (89\%) of data sources on emerging technology forecasting are related to patents, publications and News articles~\cite{viet2021data}. The rapid advancement of text-mining techniques enhances data diversity and offers new, unique insights into both explicit and latent knowledge about the nature, structure, and functions of emerging technologies. 

This paper presents a dataset of two broad components. The first comprises entity embeddings of 23k technology-adjacent entities (TA23k) with a hierarchical structure, including an identifiable subset of 100 manually verified, highly recognisable emerging technologies (ET100). The second component is a set of technology indices that can be used to filter both mature and emerging technologies from the universe of technology-adjacent entities. Those indices study the detection and diffusion of technologies ranging from old inventions to the latest advancements in various research and industry fields included in Wikipedia. This dataset aims to help researchers, policymakers, and corporations make informed decisions, allocate resources wisely, and foster the development of these technologies for sustainable growth and competitive advantage. By identifying these emerging technologies early, stakeholders can better prepare for the changes and opportunities they present, encouraging proactive adaptation and strategic planning. 

\begin{figure}[hbt!]
  % \centering
  \includegraphics[height=\textheight, keepaspectratio=true, width=\textwidth]{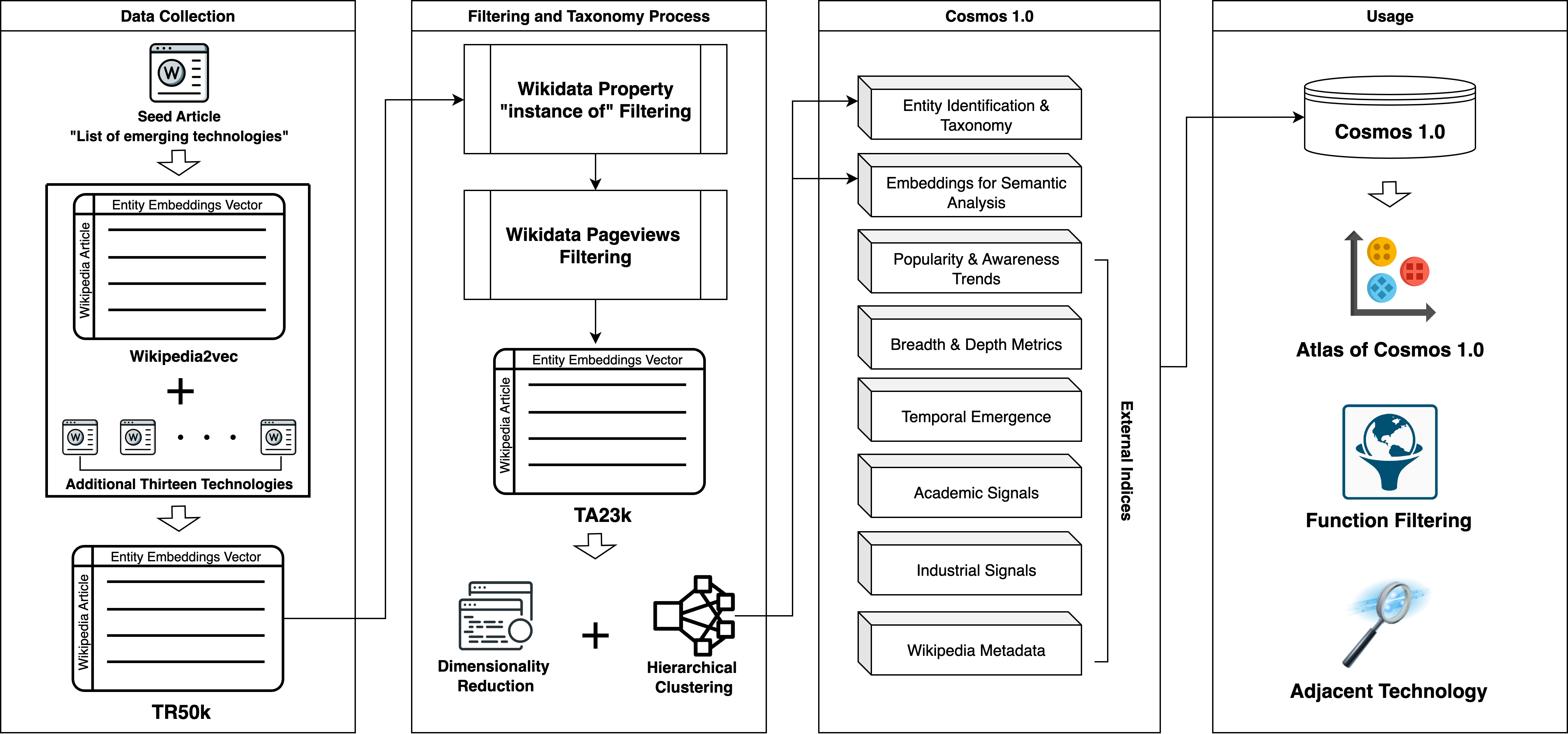}
  \caption{The repeatable workflow of creating and mapping the Cosmos 1.0 Dataset. It illustrates the simplified process of data collection, taxonomy, all features and potential usages of the Cosmos 1.0 dataset.} 
  \label{fig:fig1}
  \vspace{-4pt}
\end{figure}

This study uses a "bottom-up" approach to identify and explore the underlying structure of technology-adjacent space by leveraging the Wikipedia corpus and NLP techniques. Wikipedia is edited by numerous experts and contains texts from reliable sources~\cite{jemielniak2019wikipedia, mesgari2015sum}. For most technologies, Wikipedia provides relevant description text and links relevant articles using hyperlinks. Wikipedia2Vec, a pre-trained language model with entity embeddings, enables us to filter technology-adjacent articles related to emerging technologies based on cosine similarity~\cite{yamada2018wikipedia2vec}. Entity embeddings provide context-specific representations of real-world entities (Wikipedia article) compared to the more generalised semantic and syntactic information captured by word embeddings. After defining a universe of technology-adjacent entities from the Wikipedia corpus, dimensionality reduction algorithms and clustering algorithms are used to detect and visualise the clusters and hierarchical structure among the technology-adjacent space. The output of this "bottom-up" approach is a three-level hierarchical tree called three meta tech-clusters (TC3), seven theme tech-clusters (TC7) and ET100 from top to bottom level. The ET100 at the bottom level are the 100 emerging technologies that have been manually verified from the larger technology-adjacent universe. In the meantime, we manually name the three meta tehc-clusters and the seven theme tech-clusters based on the typical characters of the cluster of those technologies (See Subsection \hyperref[subsec:hierarchicaltree]{Hierarchical Structure of Emerging Technologies} for more details). Moreover, we collect and create a series of external indices for the final dataset (See Section \hyperref[sec:datarecords]{Data Records} for more details). The workflow of the Cosmos 1.0 dataset is shown in Figure \ref{fig:fig1}.

% The workflow of the method is shown in Figure \ref{fig: workflow}. 

\section*{Methods}
We aim to leverage large language models to automatically identify rising technologies across a broad spectrum of research areas instead of concentrating on just one area or limiting ourselves to technologies that have only recently gained popularity. Current lists of emerging technologies~\cite{gartner2024emerging30, wef2024top10, mckinsey2024techtrends} are often limited and lack depth and complex categorisation. We use insights from language models to streamline the exploration of emerging technologies and to categorise them into clusters with machine-learning methods. We also construct technology indices from reliable third-party sources to quickly and accurately position potential emerging technologies within the technology-adjacent space, validating the approach and strengthening the dataset. These are designed to offer governments, entrepreneurs, and researchers valuable perspectives. Since technologies continuously evolve, we document in detail the processes of collecting data, creating indices, and building Cosmos. It facilitates future updates to the next generation of the Cosmos dataset.

\subsection*{Data Collection based on Wikipedia2Vec}
\label{subsec:datacolle}
One significant drawback of patents and publications is that the naming of emerging technologies can vary across different publications, depending on the authors’ preferences and interpretations~\cite{jaeger2015impact}. Wikipedia’s advantage lies in providing each technology with a consistent and universally recognised name and a detailed and standardised description. Previous efforts have utilised the hyperlink structure of Wikipedia to identify more extensive lists of emerging technologies~\cite{bonaccorsi2020emerging}. However, these approaches overlooked the textual content of Wikipedia and failed to recognise the structure of these technologies or create additional metrics for comparing them. 

Wikipedia2Vec~\cite{yamada2018wikipedia2vec} is a toolkit that generates embeddings from each article and word within Wikipedia. With this toolkit, Wikipedia articles are trained as entities and thus represent the concepts and linkages within a broad context of 2.6 million other articles. The embeddings for the English language Wikipedia (2018) are used in this research. It is a powerful tool to leverage the vast amount of knowledge encoded in Wikipedia to create high-quality word and entity embeddings. It uses a skip-gram model, a type of neural network architecture that learns to predict the context of a given word based on its embedding. It then applies modifications to the original skip-gram model to incorporate the Wikipedia structure. For example, it uses the categories, links, and redirects in Wikipedia to define the context of a given word or article. It also uses a negative sampling technique to train the model efficiently. Word embeddings represent words in a vector space, while entity embeddings do the same for Wikipedia entities, such as articles for people, places, and concepts. Since the embedding space captures semantic similarities based on Wikipedia's context, co-occurrence, and linkage structure, we retrieve emerging technology-related articles using cosine similarity measurements. 

To create the Cosmos 1.0 dataset, we start with the seed article "List of emerging technologies" and collect the 100,000 most similar words and entities based on cosine similarity to the seed. The whole list contains 45,149 words and 54,851 entities. Since entity embeddings offer the advantage of capturing entities' specific contexts and relationships more accurately than word embeddings, we only keep the entities as the universe of emerging technology "candidates". Moreover, we incorporated an additional thirteen technologies found in Wikipedia that were not initially included in our set. These technologies are: "Advanced Computer Techniques", "Biopharmaceutical", "Communication", "Computer security software", "Customer relationship management", "Educational technology", "Electronic data interchange", "Enterprise resource planning", "Financial technology", "Genetics", "Microsatellite", "Online service provider", "Research". Those technologies align with other emerging technology frameworks, including World Economic Forum, OECD, and MIT Technology Review.

We refer to the complete entity list as TR50k (54,864 technology-related entities). It includes technologies, theories, organisations, people and other concepts, which are technology-related but not directly close to technologies themselves. These articles represent "noise" for our purposes, and we remove and minimise their impact using properties of Wikidata and pageviews. Firstly, we utilise the Wikidata property "instance of" to refine and clean this data. The "instance of" attribute connects an item with its broader category, denoting it as a particular instance within that category. Since we aim to remove those apparent anomalies and retain as many entities as possible, we manually select technology-adjacent instance labels and keep all Wikipedia articles without "instance of" information. Although this less rigorous screening may preserve a degree of noise, we consider this approach justified as it maximises the retention of potential emerging technology "candidates" within our dataset. There are 29,030 entities left after we filter the data based on critical labels of this attribute. The kept instance labels are available in supplementary information (Section 1 Supplementary Methods). Secondly, we collected pageviews of the TR50k for the last three years (2021 - 2023). We noted many articles with no or very few pageviews as being tangentially related to technologies rather than being technologies themselves.  We assume that valid technologies, even specialised ones, attract some degree of general attention and thus are not among the bottom 10\% percentile of the whole pageviews distribution. We note that the articles with the least attention, that is, those in the bottom 10\% percentile on the distribution tails for the last three years are those with less than 2085, 2039 and 1766 pageviews, respectively. To be more restrictive again, we manually raised the pageviews filter value to 2500 to filter the noise. In other words, we only retain entities that have received more than 2500 pageviews per year over the past three years. Finally, we end up with 23,544 technology-adjacent entities called the TA23k, which consist of the universe of emerging technology "candidates".

Applying the 2,500 pageviews threshold reshapes the log-transformed distribution of Wikipedia pageviews from approximately normal to positively skewed. This indicates that a large portion of ambiguous, non-technical entries have been effectively filtered out. The resulting skewed distribution aligns with the typical long-tailed patterns found in bibliometric and innovation datasets, where a small number of technologies attract widespread attention while the majority remain niche~\cite{clauset2009power}. Kostoff et al.~\cite{kostoff2004disruptive} are concerned that disruptive technologies are challenging to detect early because they often emerge in niche domains, outside mainstream research, and may be overlooked by conventional and strict metrics. To validate that the threshold does not exclude meaningful technologies, we compared our filtered list against established foresight sources. Over 90\% of technologies identified by the OECD fall above the 2,500 pageviews threshold (See Subsection \hyperref[subsec:oecd]{Comparison ET100 with other emerging technology lists} for more details). Together, distributional evidence and benchmarking results provide converging support for the 2,500 pageviews threshold as an empirically grounded method for curating emerging technologies, while acknowledging that additional methods (e.g., expert review) may be required to capture early-stage innovations with limited public visibility. 

Since we frequently mention technology-related entities, technology-adjacent entities, and emerging technologies in this paper, we distinguish these terms to avoid ambiguity. Technology-related entities, exemplified by TR50k, constitute the broadest set of technologies identified on Wikipedia through cosine similarity, though this set contains considerable "noise". Within our framework, we define technology-related entities as those positioned furthest from emerging technologies. Following refinement of TR50k using Wikidata and pageviews, we derived a more focused subset, TA23k, representing technology-adjacent entities that are closer to emerging technologies, yet still affected by "residual noise" due to the limitations of our filtering process. By contrast, the 100 emerging technologies (ET100) identified in this paper through a combination of tailored indices and manual screening represent the set of genuine emerging technologies (See Subsection \hyperref[subsec:hierarchicaltree]{Hierarchical Structure of Emerging Technologies} for more details).
Next, we apply machine learning algorithms to the TA23k dataset to detect the hierarchical structure of this technology-adjacent space. This provides the initial input of approximately 23k technology-adjacent entities with 100-dimensional embedding vectors. We now have a valuable approach for identifying potential emerging technologies "candidates" within the space and exploring the relationships between different technologies and research fields.

\subsection*{Hierarchical Structure of Emerging Technologies}
\label{subsec:hierarchicaltree}

Since no globally standardised classification system exists to classify technologies, we introduce unsupervised machine learning algorithms to cluster the TA23k based on their embedding vectors. Before clustering, we use the t-distributed Stochastic Neighbour Embedding (t-SNE) algorithm to map the 100-dimensional entity embeddings space to a two-dimensional space for visualisation and further clustering analysis. The t-SNE~\cite{van2008visualizing} algorithm excels at preserving local structures within high-dimensional data, making it particularly well-suited for revealing intricate cluster patterns that linear methods, such as PCA, might miss. Its non-linear approach allows it to capture complex relationships between features, offering a more nuanced view of data's inherent structure. This preprocessing step mitigates the curse of dimensionality by simplifying the data, which enables visualising and understanding complex data structures.

The second step is to utilise agglomerative hierarchical clustering (AHC)~\cite{gowda1978agglomerative} to reveal the hierarchical structure of the TA23k. It is a bottom-up clustering method where each data point starts as its cluster, and pairs of clusters are merged as one moves up the hierarchy. The process begins with calculating the similarity (or distance) between each pair of points or clusters. It then iteratively combines the closest pairs into larger clusters until all points are merged into a single cluster or a desired number of clusters is reached. This method is known for producing a dendrogram. This tree diagram illustrates the series of merges and the multi-level hierarchy of clusters, enabling detailed analysis of data grouping and structure.

\begin{figure}[hbt!]
  % \centering
  \includegraphics[height=\textheight, keepaspectratio=true, width=\textwidth]{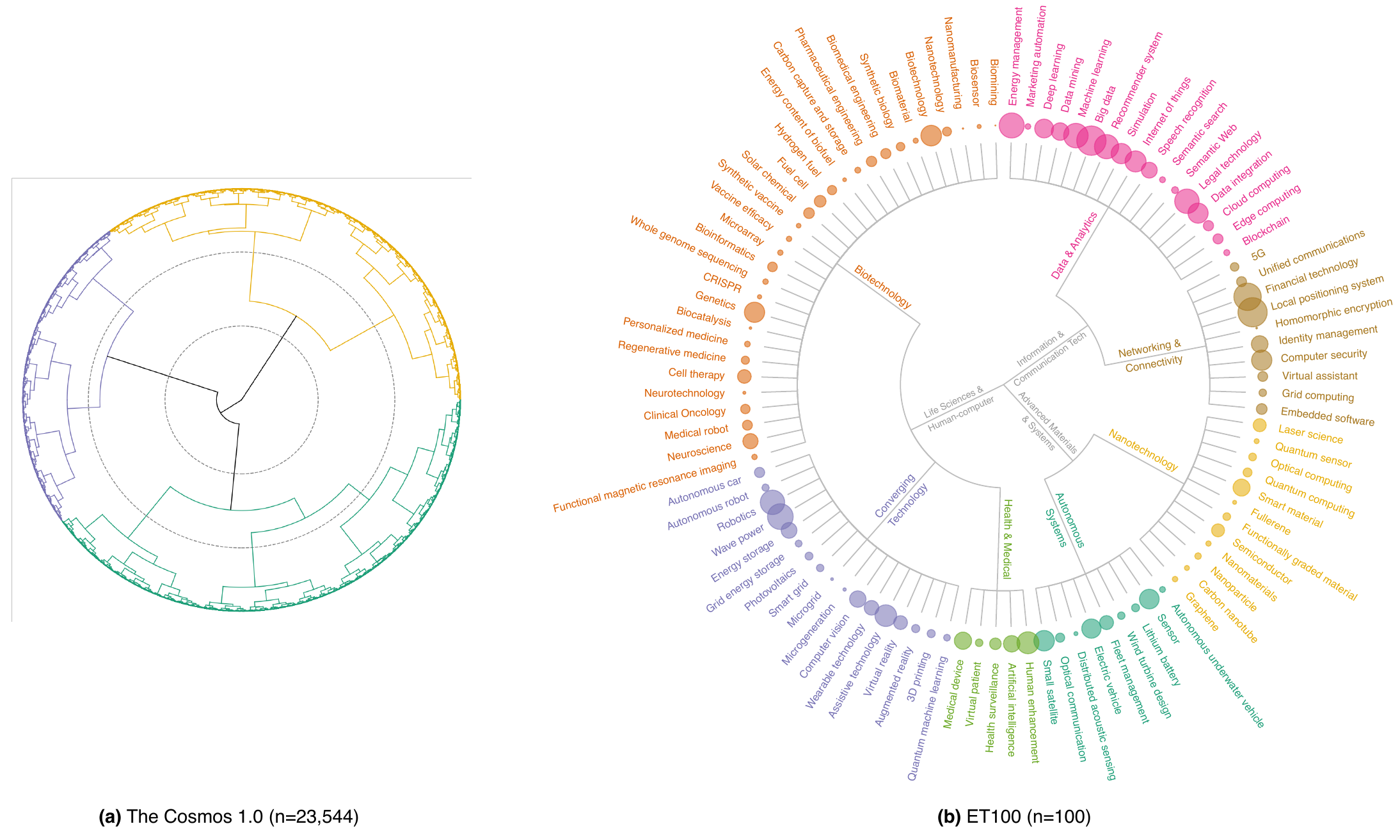}
  \caption{Radial Tree Dendorgam of Cosmos 1.0 and ET100. (a) The full radial tree dendrogram with two grey dotted circles demonstrates three and seven are optimal numbers of clusters for the TA23k. We use colours to distinguish the three meta tech-clusters (TC3). (b) The curated radial tree dendrogram shows the technologies of ET100 and cluster names of TC7 and TC3 from bottom to top. The circle sizes of ET100 represent the normalised value of a technology index called \textbf{Generality\_Index}. The theme tech-cluster "Data \& Analytics" is frequently mentioned across Wikipedia articles than other theme tech-clusters.} 
  \label{fig:fig2}
  \vspace{-4pt}
\end{figure}

As Figure \ref{fig:fig2}(a) shows, each branch represents a cluster, and the height of the branches reflects the dissimilarity between merging clusters. The dendrogram reveals two levels of optimal clustering: one at seven theme tech-clusters (TC7) and another at three meta tech-clusters (TC3). At the TC7 level, the dendrogram displays smaller, more finely grained clusters, indicating greater similarity within these groups. This level is suitable for detailed analyses with significant nuances between data points. A more pronounced gap is observed before merging into three clusters as we move higher up the dendrogram. This considerable increase in dissimilarity suggests a natural consolidation of the data into three broader categories, each representing a more general grouping of the data points. These broader clusters are ideal for high-level insights and understanding overarching patterns within the dataset. 

% The workflow of the method is shown in Figure \ref{fig: workflow}. 

Within the framework of the TC7, a curated compilation of 100 emerging technologies (ET100) was manually reviewed. For each thematic tech-cluster, we first ranked a broad pool of candidate technologies using the \textbf{Google\_Patent\_Counts\_2023} feature (See Section \hyperref[sec:datarecords]{Data Records} for more details), which captures the number of patents filed in 2023 on Google Patents that explicitly mention the technology. This patent-based signal served as a proxy for recent technological activity and intensity of innovation. Our experts manually assessed the top 1,000 technologies per cluster, using qualitative criteria such as investment potential, public and industry visibility, and relevance to broader sociotechnical trends. From this refined pool, 91 technologies were selected in the seven theme tech-clusters. Moreover, we refer to other prominent technology listings by leading global organisations such as the OECD, WEF, and Gartner, resulting in a final list of 100 emerging technologies. We ensure that each identified technology aligns precisely and unambiguously with one of the seven thematic groups with manual inspection, facilitating thematic cluster-level analyses. Figure \ref{fig:fig2}(b) provides a dendrogram that illustrates the hierarchical relationship and integration among ET100, TC7, and TC3. This structured approach enables the exploration of various levels and relationships within the hierarchy, including the dynamics between parent and child entities. 

In addition to the hierarchical structure, the ET100 gathering sequence is also worth investigating. Close-knit technologies are more likely to be grouped in the same cluster. The spectrum of technologies ranges from foundational ones, such as cloud computing and big data, to specialised innovations like CRISPR and quantum computing. Foundational technologies provide the base infrastructure, while data and analytics technologies, including machine learning and data mining, build on this foundation to process vast data volumes. Specialised industrial technologies, such as robotics and 3D printing, transform specific sectors, while cutting-edge technologies like graphene represent the forefront of scientific research. Energy and environmental technologies focus on sustainability, while health and biotechnologies, such as regenerative medicine, are critical for medical advancements. This progression reflects a move from broad, widely adopted technologies to experimental and transformative innovations at the cutting edge of technological progress. Each technology contributes to an interconnected ecosystem of advancements, pushing the boundaries of the digital and physical worlds.

The data-centric approach utilised in developing the Cosmos 1.0 stands out for its robustness and verifiability. It facilitates the analysis of connections between technologies and their relatedness. Unlike traditional methodologies that rely on manual, high-level analyses or one-off in-depth studies, our method offers comprehensive mapping capabilities compared to other emerging technology lists of OECD and the WEF, ensuring complete and precise alignment. It is designed to allow for the future inclusion or exclusion of technologies as the landscape evolves. Due to its hierarchical structure and extensive technology corpus, it also supports more detailed explorations within specific areas. This methodology ensures that both large-scale and niche technologies are included, spanning the breadth of the landscape. The utility and relevance of Cosmos 1.0 are demonstrated through the application of external technology indices, as outlined in the next section.

\subsection*{Technology Indices}
We create multiple metrics, including the Technology Awareness Index, Generality Index, Deeptech Index, Age of Tech Index and Technology Proximity Index, to capture each technology's growth, popularity, research depth and emergence in our TA23k, respectively. The "Technology Awareness Index" measures the evolving public and academic interest in emerging technologies by analysing their Wikipedia pageviews trends over time. The "Generality Index" evaluates emerging technologies' broad applicability and foundational importance by analysing their prevalence across diverse contexts within Wikipedia, providing insights into their potential as general-purpose innovations. The "Deeptech Index" leverages advanced data analytics to quantify technologies' scientific depth and innovation potential, providing a crucial tool for identifying and assessing Deeptech activities within the broader technology landscape. The "Age of Tech Index" utilises historical literature data from the extensive Google Books collection to determine the birth year of technologies, providing a foundational metric for analysing their emergence, societal impact, and evolution over time.

\subsubsection*{Technology Awareness Index}  
The use of Wikipedia pageviews to measure the growth of interest in emerging technologies is based on the assumption that increased public and professional curiosity about new technologies translates to more visits to their respective Wikipedia articles. By tracking the pageviews statistics over time, researchers can potentially gauge the rising or waning interest in specific technologies. This method offers a proxy measure of popularity or awareness rather than direct technological adoption or development. It leverages the accessibility and extensive use of Wikipedia as a primary source of information, making it a useful, albeit indirect, tool for analysing trends in technological engagement and public interest. 

To calculate the "Technology Awareness Index", we use linear regression on the pageviews data from Wikipedia articles over the past three years. By plotting the yearly pageviews against time and fitting a linear regression model, we derive the slope of this line, which represents the rate of change in pageviews. A positive slope indicates an increase in interest, reflected by more pageviews over time, whereas a negative slope suggests declining interest. This slope, quantified as the index, provides a numerical value that helps gauge the growth or decline in attention toward specific technologies, offering insights into trends in technological engagement.

\subsubsection*{Generality Index}
The aim of the "Generality Index" is to quantify and understand emerging technologies' versatility and foundational nature. It quantifies the breadth of application of a technology by analysing its presence across different contexts within Wikipedia's vast knowledge repository. This index uses computational linguistics and document frequency techniques to measure how often technology is mentioned across various Wikipedia articles. 

To construct the index, each technology is represented by a canonical search phrase (e.g., "artificial intelligence", "quantum computing"), which is queried using the Wikipedia API (https://en.wikipedia.org/w/api.php). The API returns a value indicating the number of unique Wikipedia articles in which the phrase appears (totalhits). This raw count serves as the Generality Score, a proxy for the document frequency (DF) of the term across Wikipedia. 

A higher Generality Score indicates that technology is mentioned in broader contexts, suggesting its versatility and importance as a general-purpose technology~\cite{bresnahan1995general, bekar2018general} that serves as a foundation for innovation in multiple industries and markets. In contrast, a lower score indicates a more specialised technology with limited applications. This index supports the identification of critical generative technologies that serve as essential building blocks across many industries and sectors. Such technologies are pivotal in driving continuous innovation and have a broad impact due to their wide applicability. 

The Generality Index helps researchers, policymakers, and industry leaders recognise technologies likely to be more influential and essential for future developments. Essentially, the Generality Index offers a systematic approach to gauge technologies' universality and potential ubiquity, aiding in strategic decision-making on research focus, funding allocation, and policy formulation. Limitations of this measure include coverage bias, temporal lag, and focus on popularity rather than impact. It can also be influenced by language and cultural differences and does not account for the depth or context of technology discussions, suggesting the need for third-party data for a verified assessment.

\subsubsection*{Deeptech Index}
The "Deeptech Index" is a metric designed to evaluate technologies based on their depth of scientific research and potential for disruptive innovation. Deeptech technologies are characterised by their strong foundations in science-based research and development (R\&D), often protected by intellectual property. They are distinct from technologies that primarily enhance business or service models. The index aims to distinguish these technologically and scientifically intensive technologies from those less anchored in rigorous R\&D.

We use an advanced methodology blending business and research dimensions to compute the Deeptech Index. This approach utilises the Wikipedia2Vec model, representing Wikipedia articles as embedding vectors that can be quantitatively analysed. For this index, specific Wikipedia articles related to "Business" and "Industry" serve as seeds to capture the business orientation of technology. In contrast, articles like "Basic research", "Research and development", and "Research" represent the research dimension. The cosine similarity between the embedding vectors of each technology and these seed vectors is calculated to quantify how closely a technology aligns with business or research themes.

The index calculation involves first determining the average cosine similarity of a technology to the business seeds and the research seeds. The business component is then inverted (1 - "business") to prioritise less business-oriented technologies and more grounded in deep research. The sum of the research-oriented score and the inverted business score is computed for each technology. Finally, these summed scores are ranked on a percentile scale to create the Deeptech Index, offering a relative measure of each technology's position.

This index provides a scalable and robust method for assessing Deeptech activity, overcoming limitations found in traditional datasets like patents or research investment records, which can be incomplete or outdated. By leveraging real-time and widely available Wikipedia data, the Deeptech Index offers a timely and insightful tool for analysing at scale the landscape of technologies foundational to future innovations.

\subsubsection*{Age of Tech Index}
Determining a technology's "age" is a methodological approach aimed at understanding when it becomes significant within society rather than simply tracking its first mention. This approach is encapsulated in the "Age of Tech Index", which uses a data-driven method to establish a technology's birth year based on its presence in literature, specifically leveraging the expansive Google Books database.

Google Books, a large repository of more than 40 million books in more than 400 languages, is an ideal source for this analysis. It includes collections from major research libraries like the University of Oxford Bodleian Libraries, Harvard University Library, and Stanford University Libraries. This makes it reflective of a broad societal understanding of concepts over time. This vast collection tracks the mentions of technologies from as early as 1500 up to 2019, providing a long-term perspective on technological evolution.

To determine the birth year of technology according to the Age Index, the analysis focuses on identifying the point in time when mentions of the technology reach a certain threshold, specifically, 5\% of its maximum mentions during 1900-2019. This threshold is considered significant because it indicates a notable level of recognition and integration of technology into societal or academic discourse~\cite{kousha2011assessing}. The data is analysed year by year, starting from 1900, aligning with the period of rapid technological advancement and better documentation.

This methodology is crucial for understanding technologies' emergence, growth, and relationship to other technologies. For example, it allows for examining how technologies like CRISPR and blockchain have rapidly reached broad audiences, contrasting with older technologies like neuroscience and renewable energy, which have built momentum over more extended periods. Metrics like audience reach on digital platforms provide insights into the rapid adoption of specific applications, but the Age of Tech Index offers a broader, more historical perspective. It analyses mentions across a wide range of literature, reflecting broader societal engagement with and the significance of various technologies.

\subsubsection*{Technology Proximity Index}
The introduction of the "Technology Proximity Index" aims to expedite the process of distinguishing between technologies and concepts or terms related to technology. Technology is defined as consisting of three fundamental aspects-purpose, function, and benefit-representing its timeless essence~\cite{carroll2017comprehensive}. Within the universe of Cosmos 1.0, certain companies, research institutions, concepts, and terms are highly associated with technology but do not yet meet this definition. These entities are retained to ensure the breadth and diversity of the universe. Therefore, a classifier has been trained to measure the proximity of each Wikipedia entity in Cosmos 1.0 to the concept of technology as defined.

We utilise ChatGPT with manual screening to assist in selecting entities as targets for the classifier. Specifically, we identified 100 entities from each of the seven theme tech-clusters (TC7) within Cosmos 1.0, resulting in 700 positive targets. To create a balanced input dataset, we subsequently selected 700 negative targets from the entities that were removed during the filtering process (See Subsection \hyperref[subsec:datacolle]{Data Collection based on Wikipedia2Vec} for more details). An XGBoost classifier is trained on the input data with hyperparameter tuning, resulting in an average accuracy and F1 score of approximately 86\% on the test sets. 

The classifier assigns a probability score to each entity, representing the likelihood and confidence that it aligns with the "technology." We designate this probability score as \textbf{Tech\_Proximity\_Prob}. By applying a decision threshold of 0.5, these probability scores are converted into discrete class labels, termed \textbf{Tech\_Proximity\_Index}. This process enables us to distinguish entities that closely correspond to the "technology".

\section*{Data Records}
\label{sec:datarecords}
The Cosmos 1.0 dataset is openly accessible at Figshare~\cite{cosmos_data}. The file Cosmos\_Dataset.xlsx is a table, with each row corresponding to a technology-adjacent entity with a corresponding Wikipedia page. Each row includes the following variable fields. We categorise all features into eight meaningful categories and explicitly explain their importance, providing simple examples.

\subsection*{Data Structure}
\subsubsection*{Entity Identification \& Taxonomy}
% \textbf{Entity Identification \& Taxonomy}:
These features contain identity and location in the hierarchical structure, which enables users to explore the position of a technology in the broader technology-adjacent space.
\begin{itemize}
\item \textbf{Wiki\_Entity}: a specific concept or item with a dedicated Wikipedia page. 
\item \textbf{TC3 meta tech-clusters}: Cluster labels for the three meta tech-clusters cut the dendrogram at a higher level, resulting in more generalised groups of the data points.
\item \textbf{TC7 theme tech-clusters}: Cluster labels for the seven theme tech-clusters cut the dendrogram at a lower level, yielding finer-detailed clusters that reflect closer similarities among the data points.
\item \textbf{ET100\_Flag}: Flag marks the 100 technologies manually selected.
\item \textbf{tsne\_x}: The x-axis of the 2D map in Figure \ref{fig:fig4}, also the first dimension of t-SNE on ET23k.
\item \textbf{tsne\_y}: The y-axis of the 2D map in Figure \ref{fig:fig4}, also the second dimension of t-SNE on ET23k.
\end{itemize}

\subsubsection*{Embeddings for Semantic Analysis}
% \textbf{Embeddings for Semantic Analysis}:
The feature captures the semantic context of each technology-adjacent entity, which enables clustering, similarity search, transfer learning, or visualisation in latent space.
\begin{itemize}
\item \textbf{Feature\_1-100}: Entity embedding vector representations of Wikipedia article (100 dimensions).
\end{itemize}

\subsubsection*{Popularity \& Awareness Trends}
% \textbf{Popularity \& Awareness Trends}:
This category captures public interest by tracking Wikipedia pageviews over time, enabling the identification of technologies that are rapidly gaining or losing visibility.
\begin{itemize}
\item \textbf{Pageviews\_2019-2023}: The annual sum of pageviews of Wikipedia technology-adjacent entity pages for the last five years (2019-2023).
\item \textbf{Pageviews\_3yr\_slope}: The regression slope on the annual pageviews counts of technology-adjacent entities for the last three years (2021-2023).
\item \textbf{3yr\_Awareness\_Index}: Standardised \textbf{Pageviews\_3yr\_slope} to compare how awareness of the technology-adjacent entity is growing or declining relative to others.
\item \textbf{Pageviews\_5yr\_slope}: The regression slope on the annual pageviews counts of technology-adjacent entities for the last five years (2019-2023).
\item \textbf{5yr\_Awareness\_Index}: Standardised \textbf{Pageviews\_5yr\_slope} to compare how awareness of the technology-adjacent entity is growing or declining relative to others.
\end{itemize}

\subsubsection*{Breadth \& Depth Metrics}
% \textbf{Breadth \& Depth Metrics}:
This category helps distinguish general-purpose technologies, deeptech innovations, and core technological concepts from adjacent or peripheral entities.
\begin{itemize}
\item \textbf{Generality\_Index}: Document frequency of a technology-adjacent entity mentioned across various Wikipedia articles.
\item \textbf{DeepTech\_Index}: A measure of assessing the deeptech activity of each technology-adjacent entity.
\item \textbf{Tech\_Proximity\_Prob}: Evaluate the proximity of \textbf{Wiki\_Entity} as the core technology.
\item \textbf{Tech\_Proximity\_Index}: Flag marks \textbf{Wiki\_Entity} as a core technology if its \textbf{Tech\_Proximity\_Prob} is larger than 50\%.
\end{itemize}

\subsubsection*{Temporal Emergence}
% \textbf{Temporal Emergence}:
This category helps distinguish recently emerging technologies from more established ones, supporting analyses of technology life cycles and diffusion timelines.
\begin{itemize}
\item \textbf{Age\_of\_Tech\_Index}: Predictive age of each technology-adjacent entity.
\item \textbf{First\_Pub\_Year}: The year of the first publication that mentions the technology-adjacent entity in the title, as recorded in the OpenAlex database.
\end{itemize}

\subsubsection*{Academic Signals}
% \textbf{Academic Signals}:
This category indicates how actively a technology-adjacent entity is being studied, helping to identify research frontiers and emerging academic focus areas.
\begin{itemize}
\item \textbf{Publication\_Counts\_2019-2023}: The annual counts of publications that mention the technology-adjacent entity in the title for the last five years (2019-2023) from OpenAlex.
\item \textbf{Publication\_Counts\_3yr\_slope}: The regression slope on the annual publication counts of technology-adjacent entities in OpenAlex for the last three years (2021-2023).
\item \textbf{3yr\_Publications\_Growth\_Index}: Standardised \textbf{Publication\_Counts\_3yr\_slope}.
\item \textbf{Publication\_Counts\_5yr\_slope}: The regression slope on the annual publication counts of technology-adjacent entities in OpenAlex for the last five years (2019-2023).
\item \textbf{5yr\_Publications\_Growth\_Index}: Standardised \textbf{Publication\_Counts\_5yr\_slope}.
\item \textbf{GS\_Author\_Counts}: The number of scholars, as collected from Google Scholar, who have shown interest in or are actively working in a specific technology-adjacent entity area.
\item \textbf{GS\_Author\_Counts\_Scaled}: Scale \textbf{GS\_Author\_Counts} from the smallest to the largest count, making it easier to compare across different technology-adjacent entities.
\end{itemize}

\subsubsection*{Industrial Signals}
% \textbf{Industrial Signals}:
This category helps identify technologies with growing industrial investment and innovation potential, signaling their relevance to markets and applied R\&D.
\begin{itemize}
\item \textbf{TFR}: The sum of the cumulative amount of money companies that mention specific technology-adjacent entity in their description have raised across all their funding rounds. 
\item \textbf{Google\_Patent\_Counts\_2019-2023}: The annual counts of patents that mention the technology-adjacent entity for the last five years (2019-2023) from Google Patents.
\item \textbf{Google\_Patent\_Counts\_3yr\_slope}: The regression slope on the annual Google Patent counts of technology-adjacent entities for the last three years (2021-2023).
\item \textbf{3yr\_Google\_Patent\_Growth\_Index}: Standardised \textbf{Google\_Patent\_Counts\_3yr\_slope}.
\item \textbf{Google\_Patent\_Counts\_5yr\_slope}: The regression slope on the annual Google Patent counts of technology-adjacent entities for the last five years (2019-2023).
\item \textbf{5yr\_Google\_Patent\_Growth\_Index}: Standardised \textbf{Google\_Patent\_Counts\_5yr\_slope}.
\end{itemize}

\subsubsection*{Wikipedia Metadata}
% \textbf{Wikipedia Metadata}:
This category supports qualitative analysis, global relevance assessment, and content-based filtering for downstream applications.
\begin{itemize}
\item \textbf{Wikipedia\_Content}: The full text of a Wikipedia page, excluding images and tables.
\item \textbf{Wikipedia\_Summary}: A concise overview of the main points of a Wikipedia page's content.
\item \textbf{Wikipedia\_External\_Links}: The counts of links on a Wikipedia page connect to other related Wikipedia articles, external sites, and sources that provide additional information or verify the content discussed.
\item \textbf{Wikipedia\_Num\_Ref\_Links}: The count of reference links used to substantiate the information presented in the article.
\item \textbf{Wikipedia\_Coordinates}: The geographical coordinates of locations mentioned in articles, linking to interactive maps for easy visualisation and navigation.
\item \textbf{Wikipedia\_Num\_Language}: Each technology's number of Wikipedia language editions in 2024.
\end{itemize}

\subsection*{Multidimensional Framework of the Cosmos 1.0}
Identifying emerging technologies requires multidimensional assessments rather than a single evaluation. We introduce eight categories to evaluate all technology-adjacent entities, providing a holistic framework for assessing emerging technologies within the technology-adjacent space by integrating structural, semantic, temporal, academic, industrial, and public interest dimensions.

Organising technology-adjacent entities into a hierarchical structure (TC3, TC7, ET100) provides a foundational taxonomy for understanding the emerging technology landscape (in Figure \ref{fig:fig2}(b)). Such a taxonomy helps users navigate complex domains and identify clusters of interested technologies. Katy Börner~\cite{borner2010atlas} highlights the importance of structured knowledge systems and hierarchical maps for facilitating foresight and exploration across scientific domains. This multilevel design enables both broad overviews and granular analysis of technology domains on different scales. It also allows users to quickly focus on specific domains of interest, making the dataset highly adaptable to diverse needs.

Moreover, semantic embeddings derived from Wikipedia2Vec capture rich contextual relationships among technology-adjacent entities that traditional keyword-based methods often miss~\cite{yamada2018wikipedia2vec, mikolov2013efficient}. By applying cosine similarity to rank technology-adjacent entities relative to a curated set of seeds, one can identify overlooked but semantically proximate technologies within the same cluster. It enhances the discovery of specific emerging areas within broad fields, such as "Renewable Energy Technologies" (See subsection \hyperref[subsec:cluster_similarity]{Intra-Cluster Similarity} for more details).

Prior studies~\cite{mestyan2013early, roll2016using} have demonstrated the predictive and interpretive power of Wikipedia pageviews, which can serve as a meaningful indicator of emerging attention in our dataset. Fortunato et al.~\cite{fortunato2018science} emphasise the importance of tracking temporal dynamics in scholarly and public engagement to map the evolution of scientific and technological knowledge. Accordingly, these features quantify both the volume and growth of public attention toward each technology-adjacent entity, capturing not only how widely known a technology-adjacent entity is but also how rapidly its visibility is changing. By standardising growth rates, they are helpful for comparative assessment of emerging momentum relative to other technology-adjacent entities (See subsection \hyperref[subsec:subsec:multi_indices]{Multi-Indices Filtering Strategy} for more details).

\begin{figure}[hbt!]
  % \centering
  \includegraphics[height=\textheight, keepaspectratio=true, width=\textwidth]{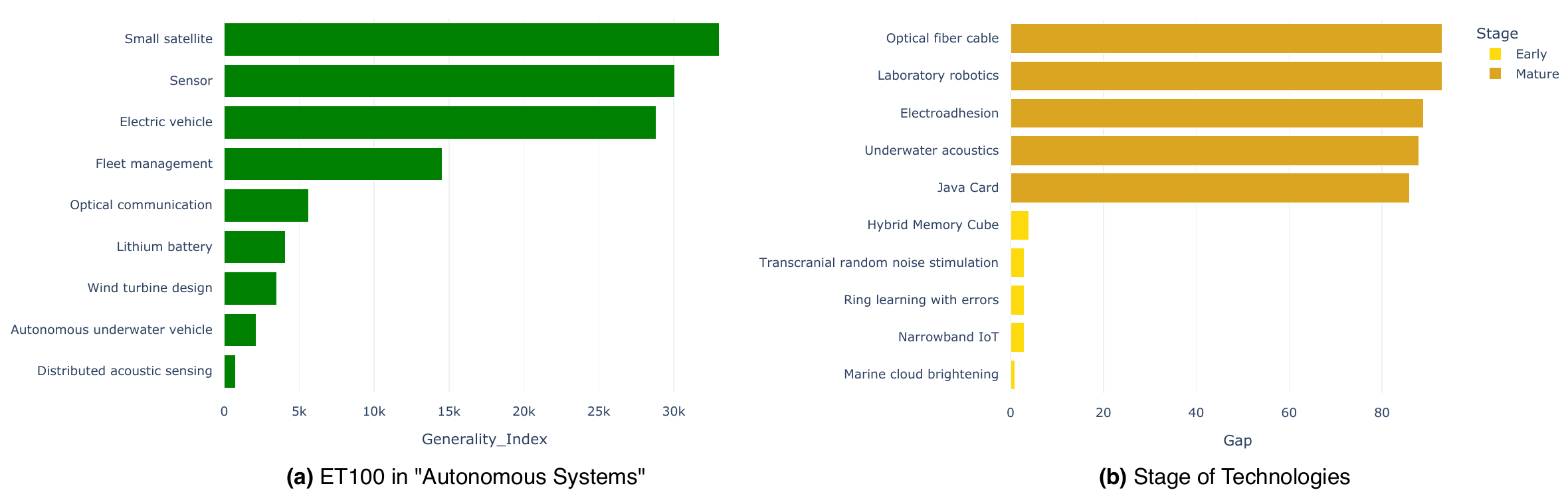}
  \caption{Example usage of features of Cosmos 1.0 (a) Technologies of ET100 in the theme tech-cluster "Autonomous Systems" are ranked by feature \textbf{Generality\_Index}. (b) Early-stage and mature-stage technologies are filtered by the difference between features \textbf{Age\_of\_Tech\_Index} and \textbf{First\_Pub\_Year}.} 
  \label{fig:fig3}
  \vspace{-4pt}
\end{figure}

Features in the Breadth \& Depth Metrics category are largely independent and exhibit low intercorrelation. Each one offers a distinct dimension and perspective on the technology universe, revealing its generality, scientific depth, and technical proximity, and is critical to assessing the potential impact and long-term development of a technology-adjacent entity. Emerging technologies often exhibit both specialisation and diffusion: some originate in narrow domains and evolve into general-purpose platforms (e.g., "Small satellite" and "Sensor"), while others gain traction through deep scientific advancement~\cite{jovanovic2005general, arora2018decline, gourevitch2021deep, chiarello2018extracting}. In Figure \ref{fig:fig3}(a), technologies with high generality indices, such as "Small satellite" and "Electric vehicle", demonstrate widespread applicability across diverse domains including aerospace, environmental monitoring, and consumer electronics. In contrast, "Distributed acoustic sensing" and "Autonomous underwater vehicle" are more specialised use cases and narrower application scopes. Traditional indicators, such as publication or patent counts, often overlook these subtleties. By integrating generality, deeptech intensity, and technical closeness, these indices offer a multidimensional view of technology-adjacent entities, enabling more accurate identification of technologies and more precise differentiation between fleeting trends and fundamental breakthroughs.

The Temporal Emergence category is crucial for identifying the stage of technologies in the innovation life cycle. Tracking temporal indicators of innovation is a well-established approach in technology forecasting and bibliometrics~\cite{callon1983translations, porter2004tech, small2009citation}. The \textbf{First\_Pub\_Year} feature captures the earliest appearance of a technology-adjacent entity in scientific literature. The \textbf{Age\_of\_Tech\_Index} estimates the onset of societal or academic recognition by identifying the first year in which the technology-adjacent entity reached 5\% of its historical peak mentions. The combination of these two features allows for distinguishing between technology-adjacent entities that have emerged recently and those that have existed longer but only recently gained traction. For example, a technology with an early publication date but a recent age index may indicate delayed adoption or a resurgence of interest. In contrast, a close alignment between the two features suggests steady and early uptake, thereby helping to classify technologies as early-stage or mature with greater precision. We focus on technologies where the \textbf{Age\_of\_Tech\_Index} exceeds the \textbf{First\_Pub\_Year}, indicating a delay between initial publication and broader recognition. To ensure relevance, we retain only those with an \textbf{Age\_of\_Tech\_Index} after 2010 and a \textbf{First\_Pub\_Year} after 1920. The size of this gap ranks the technology-adjacent entities: a larger gap suggests a mature technology-adjacent entity with delayed adoption, while a smaller gap signals a recently emerging technology-adjacent entity. From the top and bottom 60 entities, five technologies are manually selected as examples for each group shown in Figure \ref{fig:fig3}(b).

Features in the Academic Signals quantify the intensity and growth of scholarly activity, capturing how actively it is currently being studied and how rapidly that attention is growing. Increasing publication counts and growth slopes indicate an increasing interest in research~\cite{cozzens2010emerging, arora2013capturing}, while standardised indices allow fair comparison among domains with different baseline activity levels. The \textbf{GS\_Author\_Counts} feature adds another dimension by measuring the size of the engaged research community. A larger community often indicates that the technology has attracted broad, interdisciplinary attention, increasing its potential for diffusion~\cite{porter2009science}. It also suggests a stronger foundation for future innovation, collaboration, and funding, making such technologies more likely to transition from emerging research areas to impactful applications.

The Industrial Signals category helps identify technologies with commercial attraction by capturing signs of capital investment and patent data. Capital investment and patent data serve as practical indicators for identifying technologies with strong commercial potential and strategic value~\cite{ernst2003patent, ghosh2010venture, caviggioli2016technology}. For example, a high \textbf{TFR} indicates that startups referencing a technology have successfully attracted significant investment, reflecting market confidence and anticipated demand. Meanwhile, steep patent growth slopes represent intensifying innovative activity and protection efforts, which often precede product development and commercialisation. When these signals align, they help screen potential technologies that are strategically positioned for industry adoption and policy support. Therefore, this category reflects the market closeness, investment appeal and technological maturity.

The Wikipedia Metadata category provides rich textual and structural signals that enable qualitative evaluation, content filtering, and assessment of global relevance for technologies. Wikipedia content and link structures reflect collective intelligence and serve as proxies for public understanding, topical coverage, and notability~\cite{holloway2007analyzing, yasseri2012circadian}. Features such as full article content and summaries support semantic analysis and contextual enrichment, while external and reference links serve as indicators of information quality and interconnectedness, similar to how citation networks are used in scientometrics. Compared to traditional bibliographic databases, these Wikipedia-based features offer open access and real-time insights across a broader spectrum of knowledge domains. They are particularly useful for downstream tasks such as filtering ambiguous entries and identifying technologies with international presence.

\section*{Technical Validation} \label{sec:tech_val}

\subsection*{Comparison indices with third-party data sources}
Correlation analysis is used to validate the indices in our dataset. Employing three different correlation tests, including Pearson, Spearman, and Kendall Tau, provides a robust analysis by measuring relationships from various perspectives, accommodating different data characteristics like non-normal distributions and outliers, and confirming consistency across tests for more reliable results. This approach offers a statistical method to measure and establish the strength and direction of relationships between the dataset's indices and external, reputable data sources such as publication counts from OpenAlex, total funds raised from Crunchbase and patent records from Google Patents. It provides empirical evidence of validity, showing that these indices reflect real-world trends and activities surrounding technologies. By using correlation analysis in this context, we quantitatively assess the reliability of these indices as accurate indicators of technological significance and impact, thereby supporting the robustness of the dataset's construction.

There are also limitations to this approach. Correlation does not imply causation; it merely indicates patterns of association between two variables without establishing a direct cause-and-effect relationship. Additionally, the validity of the results depends heavily on the quality and relevance of the third-party data used. Since the third-party data are unavailable for all 23,544 technologies, the validation process may not comprehensively cover the entire dataset, potentially leading to biases or gaps in the validation efforts. However, all third-party data used had significant overlap in entities with the whole technology-adjacent space (TA23k), with all having more than two thousand data points in common.

Table \ref{table:1} summarises the correlations between the four constructed indices and the third-party data. To justify the Awareness Index, the growth in public interest in technologies based on Wikipedia pageviews, we evaluate the growth in academic interest in technologies based on publication counts from OpenAlex. We use the same method and compare these two trends through correlation analysis. For the trend of the last three years, both Pearson and non-parametric tests (Spearman and Kendall Tau) indicate a statistically significant positive correlation, supporting the reliability of the Awareness Index as an indicator of growing interest in these technologies over a shorter timeframe. It statistically confirms that increases in Wikipedia pageviews are associated with increases in scholarly publications. The significance of these correlations in the shorter term suggests that the Awareness Index effectively reflects real-time shifts and trends, making it a reliable tool for gauging immediate academic and public interest in new technologies. However, for the 5-year periods, only the non-parametric tests remain significantly positive, while the Pearson correlation does not show significance. This divergence may suggest that over longer periods, the relationship between public interest and scholarly output becomes less linear or direct, possibly due to the maturation of the technology or shifts in research focus. These findings highlight the importance of considering both short-term and long-term trends in technology awareness and their relation to academic activities. Granger causality could further explore the dynamic relationship between public interest and scholarly activity, helping to determine if there's a time-lagged effect where public interest in Wikipedia predicts subsequent changes in research activities. The limited data points and the requirement for data stationarity restrict the effectiveness of our analysis in this scenario. 

\begin{table*}
  \raggedright
  \caption{Technology Indices and their Correlations with Third-Party Data Sources. The statistical tests validate the accuracy, utility, and significance of the internal Cosmos 1.0 dataset by demonstrating that the internal structure of the Cosmos technology model aligns with research metrics, patent counts, and venture capital investments.}
  \label{table:1}
  \begin{tabular}{cccccc}
    \toprule
    \textbf{Technology Indices (Constructed)} & \textbf{Third-Party Data Sources} & \textbf{Sample Size} & \textbf{Pearson} & \textbf{Spearman} & \textbf{Kendall Tau} \\
    \midrule
    Awareness Index (3yr) & Publication Counts Trend (3yr)   &   19,080  &   $0.0146^{**}$ &   $0.1045^{***}$  &   $0.0725^{***}$   \\
    Awareness Index (5yr) & Publication Counts Trend (5yr)   &   19,078  &   $0.0071$ &   $0.1379^{***}$  &   $0.0956^{***}$   \\
    \midrule
    Generality Index & Patent Counts 2023   &   12,938  &   $0.2043^{***}$ &   $0.2666^{***}$  &   $0.1807^{***}$   \\
    Generality Index & Total Funds Raised   &   4,101  &   $0.2047^{***}$ &   $0.0256$  &   $0.0170$   \\
    \midrule
    Deeptech Index & Wikipedia Reference Links   &   22,109  &   $0.1305^{***}$ &   $0.1794^{***}$  &   $0.1212^{***}$   \\
    Deeptech Index & Scholar Counts   &   2,272  &   $0.0439^{**}$ &   $0.0712^{***}$  &   $0.0475^{***}$   \\
    \midrule
    Age of Tech Index & First Publication Year   &   15,614  &   $0.2897^{**}$ &   $0.4152^{***}$  &   $ 0.2942^{***}$   \\
    
    \bottomrule
\end{tabular}
     \vspace{1ex} % Adds a small vertical space
     \par
     \centering { $^{**}$ Significant at the 5\% level; $^{***}$ Significant at the 1\% level.}

  \vspace{2ex} % Adds a small vertical space
  \raggedright
  \small { \textbf{Third Party Data Sources:} Publication Counts (Microsoft Academic / OpenAlex); Patent Counts (Google Patents); Total Funds Raised (Crunchbase); Reference Links (Wikipedia); Scholar Counts (Google Scholar / League of Scholars); First Publication Year (Microsoft Academic / OpenAlex).}

\end{table*}

The correlation analysis results for the Generality Index, which assesses a technology-adjacent entity's breadth of application across various contexts within Wikipedia, reveal significant insights about its broader relevance and applicability. It can serve as an indicator of a general-purpose technology (GPT). General-purpose technologies are defined by their wide-ranging impact across many sectors, fundamentally altering industries and economies. By measuring the frequency and diversity of mentions across various contexts, the Generality Index captures an essential characteristic of GPTs: their pervasive applicability. The significant positive Pearson correlation between the Generality Index and patent counts for 2023, with a coefficient of 0.2043, indicates that technologies with a higher generality score tend to have more patent filings. This suggests that technologies mentioned frequently across different Wikipedia articles are more likely to be innovative and patentable. Similarly, the positive Pearson correlation with total funds raised (coefficient of 0.2047) underscores that more general technologies attract more investment, reflecting their potential commercial viability. However, the lack of significant positive correlations in the Spearman and Kendall Tau tests with total funds raised suggests that this relationship may not be strictly monotonic, indicating variations in how generality impacts funding across different types of technologies. The observed variation in patenting levels across industries can be meaningfully correlated with the specific characteristics of the technologies and the R\&D process, particularly in relation to the nature of the technological regime~\cite{arundel1998percentage, breschi2000technological}. Overall, these results support the validity of the Generality Index as a measure of a technology's broad relevance and potential impact in both academic and commercial areas.

The significant positive correlations across all tests for the Deeptech Index provide substantial support for its effectiveness in capturing the scientific and innovative depth of technologies. The correlation derived from Wikipedia2Vec cosine similarity measures with real-world data, such as reference links and scholar counts, indicates that the Deeptech index represents relevant aspects of advanced technologies as they are recognised and valued in practical and academic domains. This suggests that quantifying technological alignment with deep research and business aspects via Wikipedia2Vec embeddings reliably reflects a technology's depth and impact in academia. It validates using Wikipedia-based data and the Wikipedia2Vec model as robust tools for assessing technological significance, bridging the gap between theoretical conceptualisation and real-world relevance and application. These findings underscore the Deeptech Index as a valuable tool for determining technologies based on their scientific rigour and potential impact, reflecting its utility in pinpointing emerging areas likely to influence future technological landscapes.

The Age Index, identifying the emergence year of technology-adjacent entities based on historical mentions, shows significant positive correlations with the \textbf{First\_Pub\_Year} from the OpenAlex database. High correlation coefficients across all three statistical tests indicate a strong alignment between the Age Index and the earliest recorded scholarly mention of technology-adjacent entities. It demonstrates the Age Index's effectiveness in approximating the time when technologies first gained noticeable recognition and began to impact scholarly discourse. The positive results reinforce the reliability of using historical documentation frequencies to estimate the birth year of technologies, highlighting the index’s value in historical analysis of technological development.

\subsection*{Comparison ET100 with other emerging technology lists}\label{subsec:oecd}
We reviewed a number of other emerging technology lists and chose to compare our emerging technology list with the robust and extensive OECD dataset~\cite{/content/publication/sti_in_outlook-2016-en}. The list in the OECD report is derived from technology foresight exercises conducted by or for national governments of several OECD countries (Canada, Finland, Germany, United Kingdom) and the Russian Federation, as well as an exercise by the European Commission. These exercises assess promising technologies over a future 10 to 20-year horizon, pooling insights from various experts and analytical techniques. The OECD limits its focus on transformative technologies that significantly impact global socio-economic conditions and address critical challenges over the next two decades. 

The ET100 list provides a more detailed and specialised overview of emerging technologies than the OECD's broader categorisations. For instance, while the OECD lists 40 technologies in total under four general fields like Digital Technologies, Biotechnologies, Advanced Materials, and Energy plus Environment, the ET100 dives deeper into specific applications and innovations within these areas—such as edge computing in digital technologies, personalised medicine in biotechnologies, and microgeneration in energy technologies. This granularity makes the ET100 more practical for stakeholders needing precise, actionable information on current technological trends, whereas the OECD's approach is better suited for broad strategic planning and policy guidance. The ET100's emphasis on cutting-edge and specific applications and a structured hierarchical system of meta tech-clusters and theme tech-clusters offers a clear, up-to-date roadmap for navigating the rapidly evolving tech landscape.  

\begin{figure}[hbt!]
  % \centering
  \includegraphics[height=\textheight, keepaspectratio=true, width=0.9\textwidth]{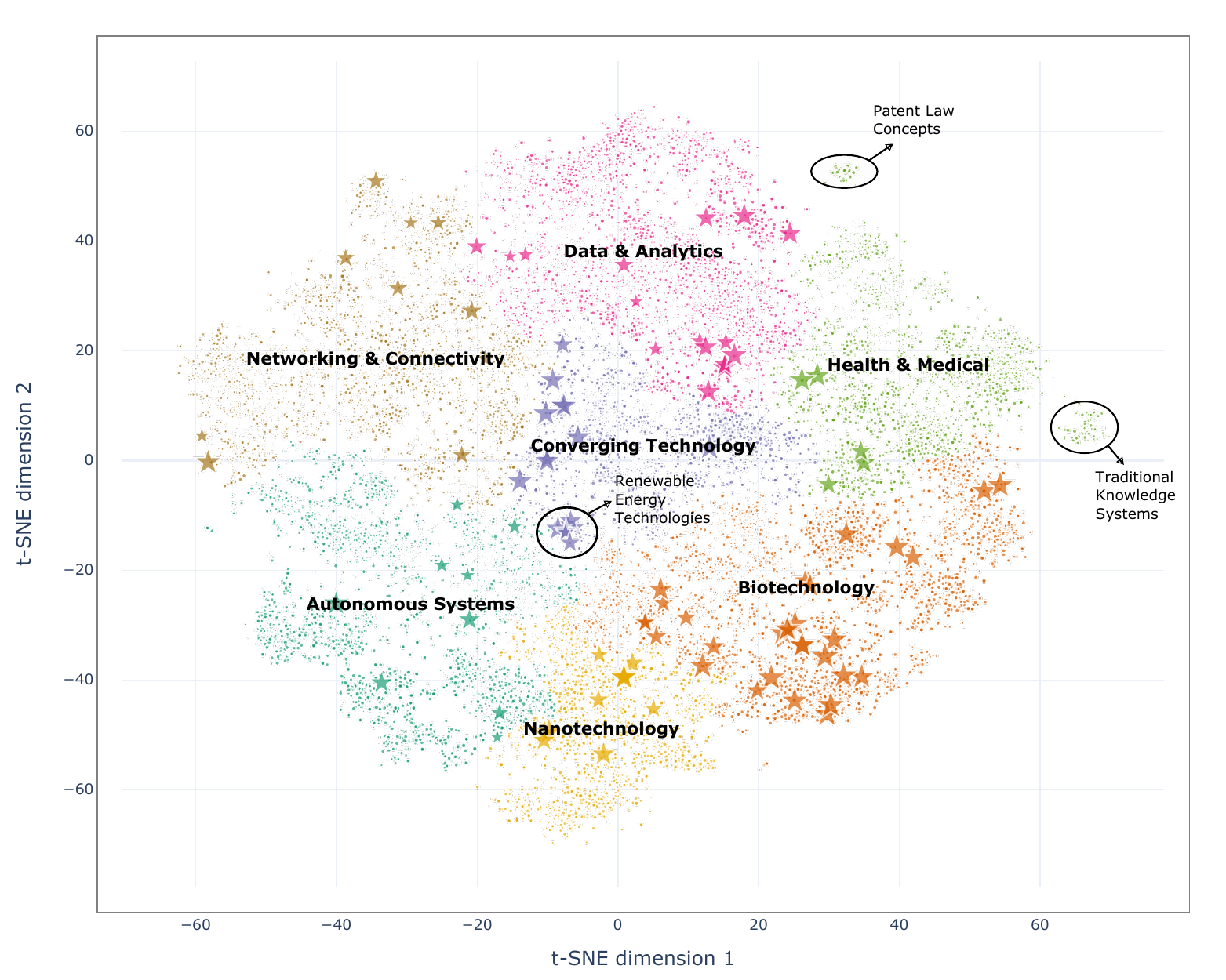}
  \caption{Map of Cosmos 1.0 (n=23,544) with highlighted ET100. Utilising the dimensionality reduction and hierarchical clustering algorithms to visualise the 7 theme tech-clusters of the TA23k in the 2D map. The dots represent 23,544 technologies in Cosmos 1.0, and the stars represent the ET100, a manual selection of widely recognised 100 technologies. This map facilitates the observation of technological distributions. The interactive version of the map at \href{https://xian-gong-elaine.github.io/Cosmos_1.0/}{here}.} 
  \label{fig:fig4}
  \vspace{-4pt}
\end{figure}

The ET100 overlaps significantly with the OECD list. The ET100 contains 31 (77.5\%) technologies that exactly match, and 37 (92.5\%) of the technologies appearing in the OECD list. The overlap primarily features in core areas like digital technologies and biotechnologies, where both identify significant trends such as AI, IoT, and synthetic biology. The ET100 and TA23k may not include some technologies due to the 2018 limitation of Wikipedia2Vec; our lists may not capture new concepts, terms, or updated information that has emerged since 2018. However, the ET100 employs a data-driven approach, using techniques (Wikipedia2Vec) and other third-party data combined with hierarchical clustering to analyse and update its list of emerging technologies. This allows for detailed, nuanced insights into specific technologies and their interconnections. In contrast, the OECD uses a more traditional method involving expert panels and foresight exercises, focusing on broader technological areas and long-term policy implications. Our approach enables more frequent updates and might also lead to a focus shift towards newer or rapidly evolving technologies, potentially omitting stable ones that still appear in the OECD's less frequently updated list. This technical and selective approach helps the ET100 maintain a more dynamic and detailed roadmap tailored to current market and innovation cycles, complementing the OECD's strategic overview. 

\section*{Usage Notes} \label{sec:usage}
The Cosmos 1.0 dataset comprises 23,544 technology-adjacent entities with a hierarchical structure, and 100 emerging technologies have been manually verified. We enhance the dataset with unique technology indices, such as Age of Tech, Deeptech, Generality, and Tech Awareness, and enrich it with metadata from Wikipedia and data from third-party sources, including Crunchbase, Google Books, and Google Scholar. In this section, we demonstrate how to utilise the Cosmos 1.0 dataset through several examples.

\subsection*{Atlas of Comos 1.0} \label{subsec:map}
We illustrate the 2D mapping of Cosmos 1.0 based on the features \textbf{tsne\_x} and \textbf{tsne\_y} from the "Embeddings for Semantic Analysis" category. Figure \ref{fig:fig4} visualises TA23k colored by seven theme tech-clusters while the stars represent the ET100. The three circular areas are the groups that are easily noticeable. The two circles at the edge are separated from the main body, forming two offshore islands (labelled "Traditional Knowledge Systems" and "Patent Law Concepts"). The third circle contains multiple technologies from ET100, notably ("Smart grid", "Microgrid", "Microgeneration", "Photovoltaics", "Grid energy storage", "Energy storage", "Wave power"), and we define these seven ET100 technologies as "Renewable Energy Technologies". We gather technologies through coordinates and notice that the two islands are technology-related terms rather than actual technologies. For example, "Traditional Knowledge Systems" primarily consists of philosophical concepts, religious doctrines, classical texts, epistemological frameworks, and metaphysical ideas from Indian, Chinese, and Buddhist traditions. "Patent Law Concepts" encompasses legal doctrines, administrative processes, classifications, and tools related to the patent system. The third circle, located at the centre of the map, includes the domains of renewable energy generation, storage, grid integration, and power system management, affirming its technological nature. This spatial layout demonstrates the utility of Cosmos 1.0 in distinguishing between technological and technology-related entities. This visual and structural differentiation is crucial for identifying emerging technologies, as it allows researchers and policymakers to focus on regions of dense innovation, observe thematic overlaps, and detect underexplored areas with high growth potential.

\begin{figure}[hbt!]
  % \centering
  \includegraphics[height=\textheight, keepaspectratio=true, width=0.9\textwidth]{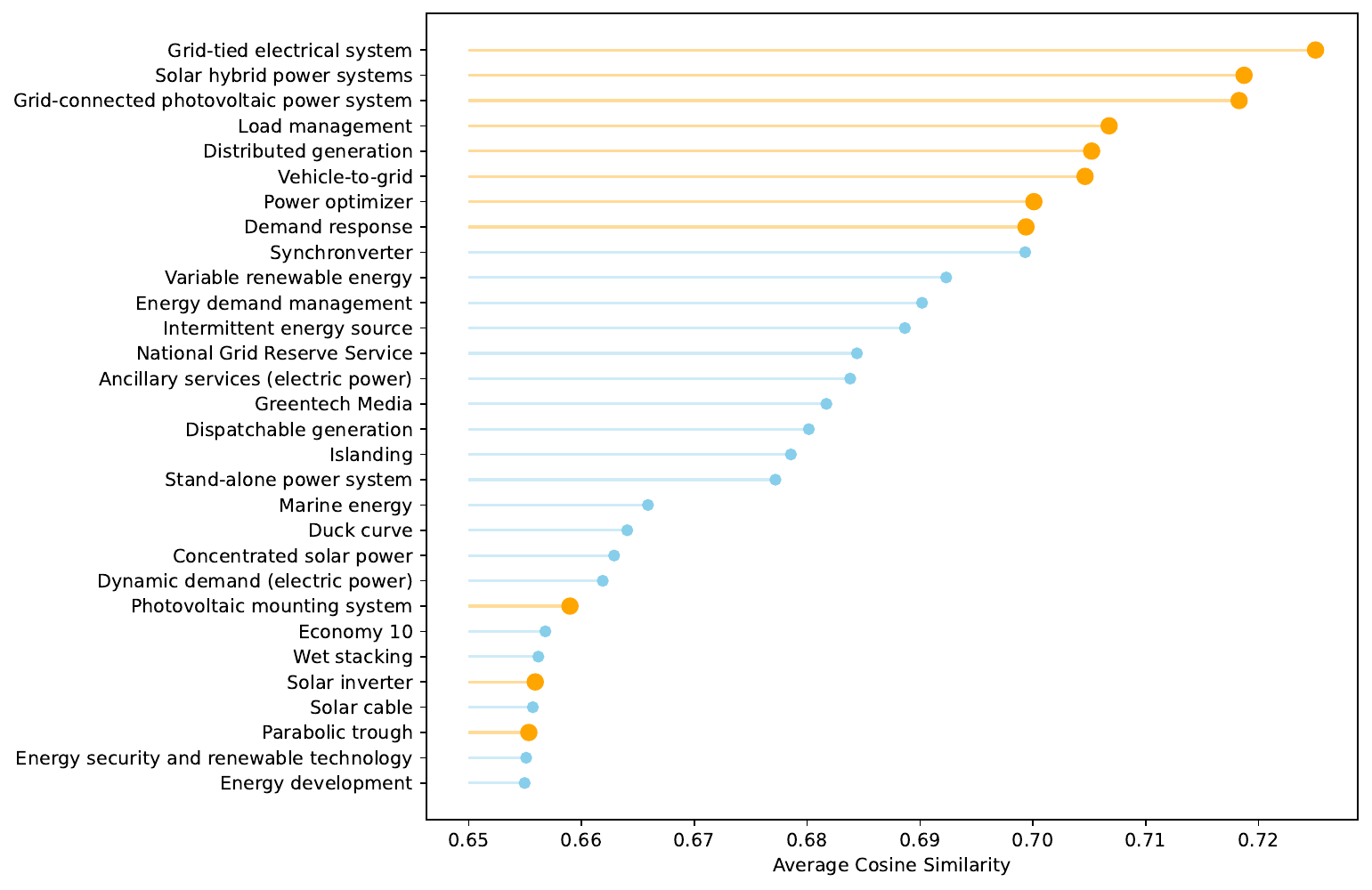}
  \caption{Top 30 technologies that are closest to the ET100 group of seven technologies "Renewable Energy Technologies" based on average cosine similarity. The higher-ranking entities demonstrate strong relevance. The entities highlighted in orange represent the technologies we observed for different potential subdomains of "Renewable Energy Technologies" in the space of solar and photovoltaic technologies.} 
  \label{fig:fig5}
  \vspace{-4pt}
\end{figure}

\begin{figure}[hbt!]
  \includegraphics[height=\textheight, keepaspectratio=true, width=0.9\textwidth]{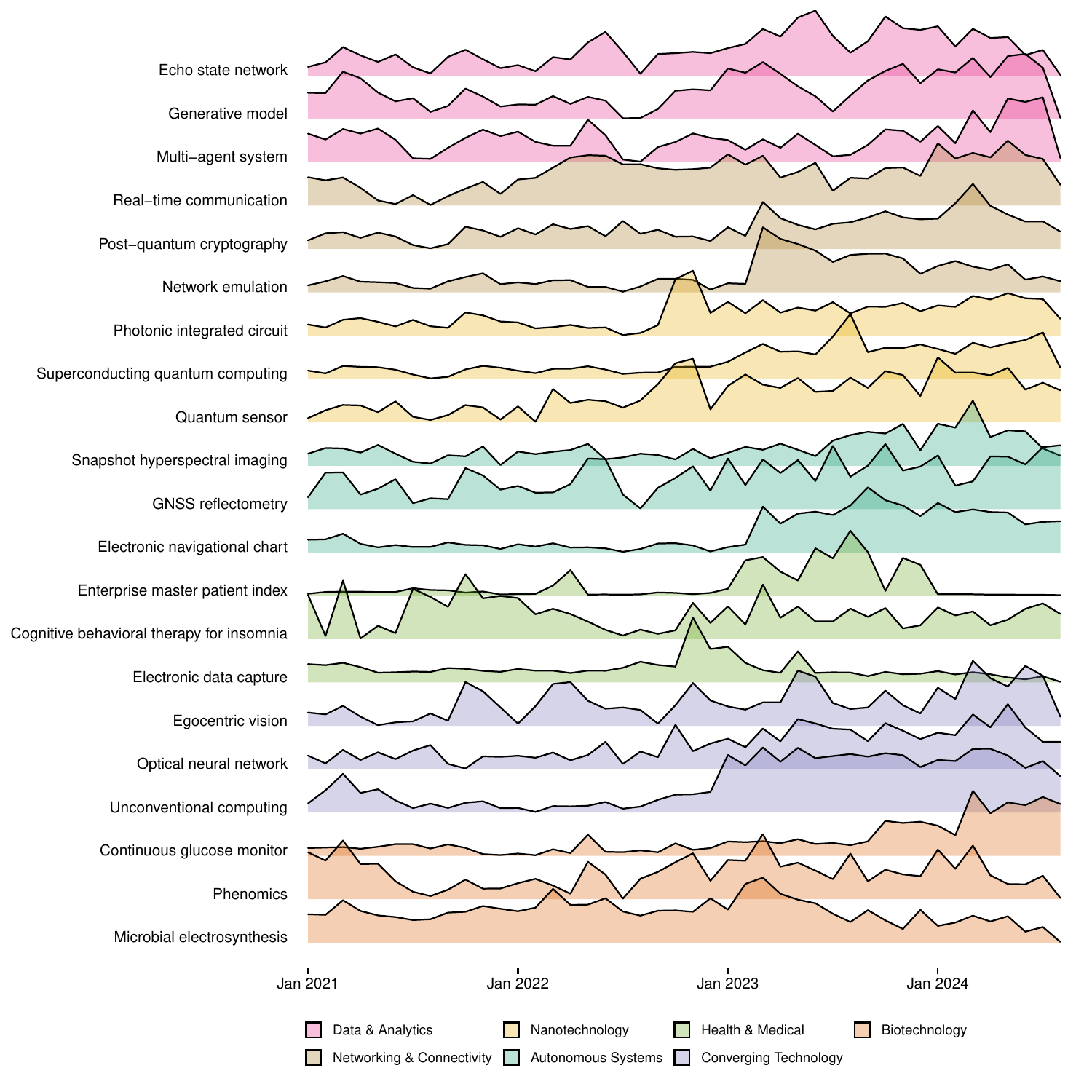}
  \caption{Monthly Pageviews Trends of Technology-Adjacent Entities of TC7. Three fast-moving frontier technologies are selected as examples for each of the seven theme tech-clusters (TC7) identified. This custom list of technologies is filtered using various technology indices that meet the different criteria. The criteria include all positive indices values, such as \textbf{Awareness\_Index}, \textbf{Publications\_Growth\_Index} and \textbf{Google\_Patent\_Growth\_Index} (See the description of indices in Section \hyperref[sec:datarecords]{Data Records}). Moreover, it only keeps the technologies that appeared after the year 1950 (\textbf{Age\_of\_Tech\_Index} > 1950).}
  \label{fig:fig6}
  \vspace{-4pt}
\end{figure}

\subsection*{Intra-Cluster Similarity}\label{subsec:cluster_similarity}
To further understand a group of closely related emerging technologies, we utilise intra-cluster cosine similarity to identify and rank technology-adjacent entities most relevant to a given seed set. We define "Renewable Energy Technologies" in the subsection \hyperref[subsec:map]{Atlas of Comos 1.0}, as the seven ET100 technologies within this circle ("Smart grid", "Microgrid", "Microgeneration", "Photovoltaics", "Grid energy storage", "Energy storage", "Wave power"). Since those technologies are included in the same theme tech-cluster "Converging Technology", we limit the scope to this theme and calculate the average cosine similarity between each candidate technology and the seven ET100 technologies within "Renewable Energy Technologies". A higher average cosine similarity indicates a stronger semantic relationship to "Renewable Energy Technologies". Based on this metric, we rank all candidate technologies in descending order and select the top 30 most relevant entities, shown in Figure \ref{fig:fig5}. The retrieved list demonstrates high quality and strong relevance. Most entities (highlighted in orange) fall within a subdomain of renewable energy, which includes solar and photovoltaic technologies (e.g., "Power optimiser", "Solar inverter", "Photovoltaic mounting system", "Parabolic trough"). Energy experts can detect more granular subdomains on the basis of the list. These ranked entities represent a cohesive technological framework encompassing generation, storage, distribution, and demand-side management. This approach helps contextualise and deepen our understanding of "Renewable Energy Technologies" and its broader ecosystem by revealing closely related and complementary technologies within the same semantic space.

\subsection*{Multi-Indices Filtering Strategy}\label{subsec:multi_indices}
Academics and researchers can leverage this dataset to track technological advancements and trends within their fields of interest using multi-index filtering. Figure \ref{fig:fig6} illustrates the monthly pageviews counts movement of an example of custom lists of technology-adjacent entities filtered by multiple technology indices. We select technology-adjacent entities from each of the TC7 theme tech-clusters that consistently exhibit strong signals across multiple features, retaining only those with positive values in \textbf{3yr\_Awareness\_Index}, \textbf{3yr\_Publications\_Growth\_Index}, and \textbf{3yr\_Google\_Patent\_Growth\_Index}. These indicators respectively capture public interest, academic momentum, and industrial innovation. Additionally, we limit the selection to technologies that emerged after 1950, based on the \textbf{Age\_of\_Tech\_Index}, to ensure relevance to recent developments. Finally, we manually review and highlight three fast-moving frontier technologies as examples of rapidly evolving technologies for each theme tech-cluster.

The multi-indices filtering strategy facilitates a deeper understanding of how specific technologies are selected and their broader implications within the technology landscape. It increases the understanding of measurements of emerging technologies. Most studies focus on the patent record~\cite{eilers2019patent, sun2024technology} or research literature~\cite{sick2022exploring}, such as robotics and internet-of-things~\cite{kose2019identifying, sun2024technology}. The Cosmos 1.0 provides a rich data source for identifying technologies on a broad scale through the eight categories.

Overall, the Cosmos 1.0 dataset provides valuable insights for researchers, policymakers, and corporations by capturing emerging technologies through complementary signals. For researchers, it enables targeted discovery of fast-evolving research frontiers. Policymakers can leverage the strategy to design policies based on technological socio-economic shifts. Corporations can use it to identify commercially promising technologies early, optimise R\&D investments, and inform strategic planning in competitive landscapes. This multidimensional framework enhances the relevance and ability of technology detection across sectors.

\section*{Data Availability} \label{sec:data}

The Cosmos 1.0 dataset generated and analysed in this study is publicly available on Figshare at \url{https://doi.org/10.6084/m9.figshare.28268561} or on Github repository at \url{https://github.com/xian-gong-elaine/Cosmos_1.0}. The file includes all the features we discussed in the paper. 

\section*{Code Availability} \label{sec:code}

The code for the analyses and the experiments in this paper is available at \url{https://github.com/xian-gong-elaine/Cosmos_1.0}.

\bibliography{references}

\section*{Acknowledgements} 
The authors extend their gratitude to Neville Stevens AO, former Chair of the NSW Government's Innovation and Productivity Council, and Darren Bell, formerly of Investment NSW, for their invaluable insights. We also thank Jessica Foote, Anthony Housego, and Neda Mirzadeh from the Australian Government's Department of Industry, whose previously commissioned works provided inspiration for this research.

\section*{Author contributions statement}
XG, CG, PXM and MAR conceived the experiments; XG conducted the experiments; All authors analysed the results, reviewed and contributed to the manuscript. CM provided insights on technology convergence; The Deeptech index was CG's idea; The Generality, Awareness and Age of Tech Indices were PXM and XG's ideas.   

\section*{Competing interests}
The authors declare no competing interests.

\end{document}